\documentclass[aps,prd,twocolumn,groupedaddress,showpacs,nofootinbib,amssymb]{revtex4}
\usepackage[T1]{fontenc}
\usepackage[latin1]{inputenc}
\usepackage{graphicx}
\usepackage[english]{babel}
\usepackage{amsmath}
\usepackage{amssymb}
\usepackage{amsfonts}

\begin{document}
\def\pp{{\, \mid \hskip -1.5mm =}}
\def\cL{{\cal L}}
\def\beq{\begin{equation}}
\def\eneq{\end{equation}}
\def\bea{\begin{eqnarray}}
\def\enea{\end{eqnarray}}
\def\tr{{\rm tr}\, }
\def\nn{\nonumber \\}
\def\e{{\rm e}}

\title{\textbf{$f(R)$ scalar-tensor cosmology by Noether symmetry}}

\author{F. Darabi$^\bullet$$^\ast$, S. Asgharinya$^\bullet$}

\affiliation{\it $^\bullet$Department of Physics, Azarbaijan Shahid Madani University , Tabriz 53741-161, Iran\\
$^\ast$Research Institute for Astronomy and Astrophysics of Maragha (RIAAM), Maragha 55134-441, Iran}

\date{\today}

\begin{abstract}
In the framework of $f(R)$ scalar-tensor cosmology, we use the Noether symmetry approach to find the cosmological models consistent with the Noether symmetry.
We obtain the functions $f(R)$ and $H(a)$, or the corresponding differential equations, according to specific choices for the scalar field potential $V(\phi)$, the Brans-Dicke function $\omega(\phi)$, some  cosmological parameters,
and the constants of motion. 

\end{abstract}
\maketitle
\section{introduction}
\label{1}
It is known that the expansion of universe is currently undergoing a period of acceleration which is directly measured by observations such as
Type Ia Supernovae~\cite{SN1}, large scale structure~\cite{LSS},
cosmic microwave background (CMB) radiation~\cite{WMAP, Komatsu:2010fb},
weak lensing~\cite{WL}, and baryon acoustic oscillations~\cite{BAO}.
There are two common approaches to explain the current acceleration
of the universe: The first one is to introduce some new cosmological components of energy sources contributing to the so called ``dark energy'' in the framework of general relativity (for a review on dark energy, see, e.g.,~\cite{Copeland:2006wr, Li:2011sd}). The second one is to generalize $R$ (Ricci scalar) gravity to some modified gravities \cite{mauro, farasot, libroSaFe}. One of the most common modified gravities is $f(R)$ gravity~\cite{f(R)-cosmo,
Review-N-O, Sotiriou:2008rp, DeFelice:2010aj, Clifton:2011jh}.
This theory relaxes the hypothesis that gravitational Lagrangian has to be a linear function of $R$, and as a minimal extension introduces an effective action containing a generic $f(R)$ function. 

On the other hand, generalized actions of a scalar field nonminimally coupled to $R$ gravity, as a generalization of Brans-Dicke theory \cite{brans}, have been extensively studied \cite{cimall}. In the present paper, we intend to more generalize such theories to include $f(R)$ gravity with a scalar field nonminimally coupled to $f(R)$ gravity. Explicitly, we aim to obtain the forms of $f(R)$, appearing in such modified action, by demanding that the Lagrangian admits the desired Noether symmetry \cite{8,10} (for a study of the Noether symmetry in various cosmological models see \cite{9}). We shall see that by demanding the Noether symmetry, we can either obtain the explicit forms of the function $f(R)$ or at least find the differential equations which can be solved to obtain $f(R)$.

In Sec. (\ref{2}) we introduce the action of a $f(R)$ scalar-tensor theory
and obtain the corresponding field equations. In Sec. (\ref{3}), we introduce in general the Noether symmetry approach, and in Sec. (\ref{4}) we apply it to the $f(R)$ scalar-tensor cosmology. In Sec. (\ref{5}), we obtain the forms of $f(R)$ or the differential equations for $f(R)$. Conclusions are given in Sec. (\ref{7}).
\section{Cosmology from scalar-tensor theories}
\label{2}

Let us consider the general action
\beq
\mathcal{A}=\int d^4x\sqrt{-g}\left(\phi^2f(R)+4\omega(\phi)g^{\mu\nu}\nabla_\mu\phi\nabla_\nu\phi-V(\phi)\right),
\eneq
where the scalar field $\phi$ is nonminimally coupled to $f(R)$, and $\omega(\phi)$ and $V(\phi)$ are respectively the Brans-Dicke parameter and the potential as generic functions of $\phi$. In order to derive the cosmological equations in a FRW metric \cite{defelice}, one can define a canonical Lagrangian $\mathcal{L}=\mathcal{L}(a,\dot{a},R,\dot{R},\phi, \dot{\phi})$, where $\mathcal{Q}=\left\{a,R,\phi \right\}$ is the configuration space and $\mathcal{TQ}=\left\{a,\dot{a},R,\dot{R},\phi,\dot{\phi}\right\}$ is the related tangent bundle on which $\mathcal{L}$ is defined, where a dot denotes derivative with respect to the cosmic time $\textit{t}$.
The variable $a$ is the scale factor in FRW metric, and all dynamical variables
$a, R$, and $\phi$ are assumed to depend just on $\textit{t}$ to restore homogeneity and isotropy. The presence of Ricci scalar in the Lagrangian needs explanation.
In fact, it is assumed that $R$, as well as $a$ and $\phi$, is a canonical variable because it is generally used in canonical quantization of higher order gravitational theories. However, such a position seems arbitrary, since $R$ is not independent of $a$ and $\dot{a}$. Hence, one can use the method of Lagrange multipliers to set $R$ as a constraint of the dynamics 
\bea
\mathcal{A}&=& \int dta^3\left\{\phi^2f(R)+4\dot{\phi}^2\omega(\phi)-V(\phi)\right. \nonumber \\    
 &+& \left.\lambda\left[R-6\left(\frac{\ddot{a}}{a}+\left(\frac{\dot{a}}{a}\right)^2+\frac{kc^2}{a^2}\right)\right]\right\}, \enea
where $\lambda$ here is a Lagrange multiplier. The variation of action with respect to $R$ gives $\lambda=-\phi^2f_R$
where $f_R :=\frac{d}{dR}$. Therefore, the above action can be rewritten as
\bea
\mathcal{A}&=& \int dta^3\left\{\phi^2f(R)+4\dot{\phi}^2\omega(\phi)-V(\phi)\right. \nonumber \\    
 &-& \left.\phi^2f_R\left[R-6\left(\frac{\ddot{a}}{a}+\left(\frac{\dot{a}}{a}\right)^2+\frac{kc^2}{a^2}\right)\right]\right\}. \enea
By integrating by parts, and neglecting a pure divergence we obtain the point-like FRW Lagrangian
\bea
\mathcal{L}&=&a^3\phi^2\left(f(R)-Rf_R\right)-6\phi^2a\dot{a}^2f_R \nonumber \\ \label{4.4}
&-& 12a^2\dot{a}\phi\dot{\phi}f_R-6\phi^2a^2\dot{R}\dot{a}f_{RR}+6\phi^2kaf_R \nonumber \\
&+& a^3\left[4\dot{\phi}^2\omega(\phi)-V(\phi)\right].
\enea
The equations of motion for $a$, $R$ and $\phi$ are obtained respectively
\bea
2f_{RRR}\phi^2a^2\dot{R}^2&+&2f_{RR}\left[2a\ddot{a}\phi^2+2\phi\dot{\phi}a^2\dot{R}\right.\nonumber \\
+\left.2\phi^2a^2\ddot{R}\right]&+&2f_R\left[4a\dot{a}\phi\dot{\phi}+\phi^2\dot{a}^2+2a^2\dot{\phi}^2+2a^2\phi\ddot{\phi}\right. \nonumber \\
&+&k\phi^2-\left.(1/2)a^2\phi^2R\right]+a^2\phi^2f(R) \nonumber \\
&+&a^2\left[4\dot{\phi}^2\omega(\phi)-V(\phi)\right]=0,
\enea
\beq
R-6\left[\frac{\ddot{a}}{a}+\left(\frac{\dot{a}}{a}\right)^2+\frac{kc^2}{a^2}\right]=0,
\eneq
\bea
Af_R\left[\phi Ra^2-6\dot{a}^2\phi-6a\ddot{a}\phi-6k\phi\right]-a^3\phi f(R) \nonumber \\
+4a^3\ddot{\phi}\omega(\phi)+12a^2\dot{a}\dot{\phi}\omega(\phi)+2a^3\dot{\phi}^2d\omega/d\phi \nonumber \\
+(1/2)a^3dV/d\phi=0.
\enea
Finally the total energy $E_\mathcal{L}$, corresponding to the $(00)$ Einstein equation is obtained as
\bea
6\phi^2a\dot{a}\dot{R}f_{RR}&+&f_R\left[6\phi^2\dot{a}^2+12a\phi\dot{a}\dot{\phi}-a^2\phi^2R\right. \nonumber \\
+ \left.6k\phi^2\right]&+&a^2\phi^2f(R)-8a^2\dot{\phi}^2\omega(\phi) \nonumber \\
&+&a^2\left[4\dot{\phi}^2\omega(\phi)-V(\phi)\right]=0.
\enea 
\section{Noether symmetry approach}
\label{3}

Let $\mathcal{L}(q^i,\dot{q}^i)$ be a canonical, non degenerate point-like Lagrangian subject to
\beq
\frac{\partial\mathcal{L}}{\partial t}=0, \hspace{1.5cm}  det H_{ij}\equiv \left\|\frac{\partial^2\mathcal{L}}{\partial \dot{q}^i\partial \dot{q}^j}\right\|\neq0,
\eneq
where $H_{ij}$ is the Hessian matrix and a dot denotes derivative with respect to the cosmic time $\textit{t}$. The Lagrangian $\mathcal{L}$ is generally
of the form
\beq
\mathcal{L}=T(\textbf{q},\dot{\textbf{q}})-V(\textbf{q}),                   \label{4.39}
\eneq
where \textit{T} and \textit{V} are the `kinetic energy' (with positive definite quadratic form) and `potential energy' respectively. The energy function associated with $\mathcal{L}$ is defined 
\beq
E_\mathcal{L}\equiv\frac{\partial\mathcal{L}}{\partial \dot{q}^i}-\mathcal{L},
\eneq
which is the total energy $T + V$ as a constant of motion. Since our cosmological problem has a finite number of degrees of freedom, we consider only point transformations.

Any invertible transformation of the generalized positions $Q^i=Q^i(\textbf{q})$ induces a transformation of the generalized velocities 
\beq
\dot{Q}^i(\textbf{q})=\frac{\partial Q^i}{\partial q^j}\dot{q}^j,   \label{4.23}
\eneq
where the matrix $\mathcal{J}=\left\|\partial Q^i/\partial q^j\right\|$ is the Jacobian of the transformation, and it is assumed to be non-zero. On the other hand, an infinitesimal point transformation is represented by a generic vector field on $Q$
\beq
\textbf{X}=\alpha^i(\textbf{q})\frac{\partial}{\partial q^i}.
\eneq
\\ The induced transformation (\ref{4.23}) is then represented by
\beq
\textbf{X}^c=\alpha^i\frac{\partial}{\partial q^i}+\left(\frac{d}{dt}\alpha^i\right)\frac{\partial}{\partial \dot{q}^j}.   \label{4.24}
\eneq
The Lagrangian $\mathcal{L}(\textbf{q}, \dot{\textbf{q}})$ is invariant under the transformation by \textbf{X} provided that
\beq
L_X\mathcal{L}\equiv\alpha^i\frac{\partial \mathcal{L}}{\partial q^i}+\left(\frac{d}{dt}\alpha^i\right)\frac{\partial}{\partial \dot{q}^j}\mathcal{L}=0,
\eneq
where $L_X\mathcal{L}$ is the Lie derivative of ${\mathcal{L}}$.
Let us now consider the Lagrangian $\mathcal{L}$ and its Euler-Lagrange equations
\beq
\frac{d}{dt}\frac{\partial\mathcal{L}}{\partial \dot{q}^j}-\frac{\partial\mathcal{L}}{\partial q^j}=0.                \label{4.25}
\eneq
Contracting (\ref{4.25}) with $\alpha^i$ gives
\beq
\alpha^j\left(\frac{d}{dt}\frac{\partial\mathcal{L}}{\partial \dot{q}^j}\right)=\alpha^j\left(\frac{\partial\mathcal{L}}{\partial q^j}\right).     \label{4.26}      
\eneq
On the other hand, we can write
\beq \label{4.26'}
\frac{d}{dt}\left(\alpha^j\frac{\partial\mathcal{L}}{\partial \dot{q}^j}\right)=\alpha^j\left(\frac{d}{dt}\frac{\partial\mathcal{L}}{\partial \dot{q}^j}\right)+\left(\frac{d\alpha^j}{dt}\right)\frac{\partial\mathcal{L}}{\partial \dot{q}^j},
\eneq
in which the first term in the RHS can be replaced by the RHS of (\ref{4.26}),
hence (\ref{4.26'}) results in
\beq
\frac{d}{dt}\left(\alpha^j\frac{\partial\mathcal{L}}{\partial \dot{q}^j}\right)=L_X\mathcal{L}.
\eneq
The immediate consequence of this result is the \textit{Noether theorem} which states: if $L_X\mathcal{L}=0$, then the function
\beq
\Sigma_0=\alpha^k\frac{\partial\mathcal{L}}{\partial \dot{q}^k},         \label{4.27}
\eneq
is a constant of motion. 

\section{Noether symmetries in scalar-tensor cosmology}
\label{4}
Considering the $f(R)$ scalar-tensor cosmology, the vector field associated
with the Noether symmetry is  
\beq
\textbf{X}=A\frac{\partial}{\partial a}+B\frac{\partial}{\partial \phi}+C\frac{\partial}{\partial R}+\dot{A}\frac{\partial}{\partial \dot{a}}+\dot{B}\frac{\partial}{\partial \dot{\phi}}+\dot{C}\frac{\partial}{\partial \dot{R}}.           \label{4.28}
\eneq
Now, the Noether symmetry exists if at least one of the functions $A$, $B$ or $C$ in the equation (\ref{4.28}) is different from zero. To investigate the existence of Noether symmetry, we should write down the equation $L_X\mathcal{L}=0$ as the following system of differential equations
\beq
3A\omega(\phi)+Ba\frac{d\omega}{d\phi}-3f_R\phi\partial_\phi A+2\omega(\phi)a\partial_\phi B=0,   \label{0.1}
\eneq
\bea
f_R\left(A\phi+2Ba+2a\phi\partial_a A+2a^2\partial_a B\right) \nonumber \\                   \label{0.2}
+a\phi f_{RR}\left(C+a\partial_a C\right)=0,
\enea
\bea
f_R\left(2A\phi+Ba+a\phi\partial_a A+\phi^2\partial_\phi A+a\phi\partial_\phi B\right)  \nonumber \\
-\frac{2}{3}\omega(\phi)a^2\partial_a B+a\phi f_{RR}\left(C+\frac{\phi}{2}\partial_\phi C\right)=0,                 \label{0.3}
\enea
\bea
f_R\left(2\phi\partial_R A+2a\partial_R B \right)+2 f_{RR}\left(aB+A\phi\right)  \nonumber \\                   \label{0.4}
+a\phi\left(Cf_{RRR}+\partial_a Af_{RR}+\partial_R Cf_{RR}\right)=0,
\enea
\beq
\phi^2f_{RR}\partial_\phi A-\frac{4}{3}\omega(\phi)a\partial_R B+2\phi f_R\partial_R A=0,                     \label{0.5}
\eneq
\beq
\partial_R A=0,                       \label{0.6}
\eneq
which are obtained by setting to zero the coefficients of the terms $\dot{a}^2$, $\dot{R}^2$, $\dot{\phi}^2$, $\dot{a}\dot{R}$, $\dot{a}\dot{\phi}$, $\dot{R}\dot{\phi}$ in $L_X\mathcal{L}=0$. Finally, we have to satisfy the constraint
\bea\label{0.8}
6k\phi^2Af_R&+&3a^2\phi^2A\left(f-Rf_R\right)-3a^2V(\phi)A \nonumber \\
&+&2a^3\phi B\left(f-Rf_R\right)-Rf_{RR}a^3\phi^2C  \\
&-&Ba^3\frac{dV}{d\phi}+12ka\phi f_R B+6ka\phi^2f_{RR}C=0. \nonumber
\enea
A solution of (\ref{0.1})-(\ref{0.6}) exists if explicit forms of $A$, $B$ and $C$ are found. By using Eq. (\ref{0.6}), the equation (\ref{0.4}) becomes
\bea
f_R\left(2a\partial_R B\right)&+&f_{RR}\left(2A\phi+2aB+a\phi\partial_a A\right) \nonumber \\
&+&a\phi\partial_R\left(Cf_{RR}\right)=0,
\enea
which can be rewritten as
\beq
\partial_R\left(2aBf_R+a\phi C f_{RR}\right)+f_{RR}\left(2A\phi+a\phi\partial_a A\right)=0,
\eneq
and solved with respect to $R$ as
\beq
2Bf_R+\phi C f_{RR}=-\left(2\frac{A\phi}{a}+\phi\partial_a A\right)f_{R}+h(a,\phi),          \label{0.7}
\eneq
where $h(a,\phi)$ is the integration constant. 
From Eq.(\ref{0.7}) we get $C$
\beq
C=\frac{1}{\phi}\left[-\left(2B+2\frac{A\phi}{a}+\phi\partial_a A\right)\frac{f_{R}}{f_{RR}}+\frac{h(a,\phi)}{f_{RR}}\right].
\eneq
Inserting $C$ into Eq. (\ref{0.2}) results in
\beq
f_R\left(A\phi-a\phi\partial_a A-\phi a^2\frac{\partial^2 A}{\partial a^2}\right)+a\left(h+a\partial_a h\right)=0,
\eneq
which is solved for $A$ and $h$ as
\beq
A=\left(c_1a+\frac{c_2}{a}\right)g(\phi) \hspace{0.6cm} and \hspace{0.6cm} h=\frac{\bar{c}}{a}\lambda(\phi),
\eneq
where $c_1$, $c_2$ and $\bar{c}$ are integration constants, and $g(\phi)$
and $\lambda(\phi)$ are some generic functions of $\phi$.
Substituting $A$ and $h$ into $C$ we obtain
\beq\label{C}
C=-\left[\frac{2B}{\phi}+\left(3c_1+\frac{c_2}{a^2}\right)g(\phi)\right]\frac{f_R}{f_{RR}}+\frac{\bar{c}\lambda(\phi)}{a\phi f_{RR}}.
\eneq
We leave the constraint (\ref{0.8}) as an equation to choose suitable potential
$V(\phi)$ and $f_R$. The remaining equations governing $B$, $g(\phi)$, $\omega(\phi)$ and $\lambda(\phi)$ are
\bea
&&3\left(c_1a+\frac{c_2}{a}\right)g(\phi)\omega(\phi)+aB\frac{d\omega}{d\phi}+2a\omega(\phi)\partial_\phi B  \nonumber \\
&-&3f_R\phi\left(c_1a+\frac{c_2}{a}\right)\frac{dg}{d\phi}=0,         \label{0.9}
\enea
\bea
&&\frac{1}{2}f_R\left(-c_1a+\frac{c_2}{a}\right)\phi^2\frac{dg}{d\phi}+\frac{\bar{c}}{2}\left(\lambda+\phi\frac{d\lambda}{d\phi}\right)\nonumber \\
&-&\frac{2}{3}\omega(\phi)a^2\partial_a B=0,        \label{1.0}
\enea
\beq
\phi^2f_{RR}\left(c_1a+\frac{c_2}{a}\right)\frac{dg}{d\phi}-\frac{4}{3}\omega(\phi)a\partial_R B=0.    \label{1.1}
\eneq
By taking $\lambda(\phi)=\lambda_0 \phi^{-1}$ in Eq. ({\ref{1.0}}) the term proportional to $\bar{c}$ vanishes. The resulting equations ({\ref{0.9}}), ({\ref{1.0}}) and ({\ref{1.1}}) are just consistent for constants $B(a,R,\phi)=B_0$ and $g(\phi)=g_0$. Equations ({\ref{1.0}}) and ({\ref{1.1}}) are satisfied
by these solutions and Eq. ({\ref{0.9}}) becomes
\beq
3\left(c_1a+\frac{c_2}{a}\right)g_0\omega(\phi)+aB_0\frac{d\omega}{d\phi}=0.         \label{1.2}
\eneq
One may set $c_2=0$ to convert this equation into a simple differential equation
\beq
\frac{d\omega}{d\phi}+\kappa^2\omega=0,         \label{1.3}
\eneq
where $\kappa^2=3c_1g_0/B_0$.\\
For the choices of the constants $c_1, g_0$ and $B_0$ resulting in $\kappa^2>0$
we have the oscillating solutions 
\beq
\omega(\phi)=\omega_0\exp({\pm i \kappa \phi}),         \label{1.4}
\eneq
whereas for the choices leading to  $\kappa^2<0$, we obtain exponential solutions
\beq
\omega(\phi)=\omega_0\exp({\pm \kappa \phi}).         \label{1.5}
\eneq
Therefore, we find
\bea\label{1.66}
A&=&c_1 g_0 a,
\nonumber \\
B&=&B_0,\\        
C&=&-\left[\frac{2B_0}{\phi}+3c_1g_0\right]\frac{f_R}{f_{RR}}+\frac{\bar{c}\lambda_0}{a\phi^2 f_{RR}}.\nonumber
\enea
\\
The existence of non zero quantities $A, B$ and $C$ accounts for the Noether
symmetry provided that $A, B, C$, $f_R$ and $V(\phi)$
satisfy the constraint ({\ref{0.8}}). This equation may be converted to a
differential equation for $f(R)$ as follows
\bea \label{0.8'}
f_R&=&\frac{1}{12k a\phi^2 c_1 g}[3a^3\phi^2c_1 g_0+2a^3\phi B]f\\ \nonumber
&-&\left[3a^3V(\phi)c_1 g_0+\bar{c}\lambda_0(Ra^2-6k)+Ba^3\frac{dV}{d\phi}\right].
\enea
We may find the constant of motion, namely the Noether charge as
\bea
\Theta_0&=&A\frac{\partial\mathcal{L}}{\partial\dot{a}}+B\frac{\partial\mathcal{L}}{\partial\dot{\phi}}+
C\frac{\partial\mathcal{L}}{\partial\dot{R}}\\ \nonumber
&=&-6\phi a^2 c_1 g_0
[2f_R (\phi a\dot{)}+\phi a\dot{R}f_{RR}]
\nonumber \\  
&-&12 B_0a^2\dot{a}\phi f_R +8B_0a^3\dot{\phi}\omega(\phi)
\nonumber \\
&+&6\phi^2a^2\dot{a}\left[\frac{2B_0}{\phi}+3c_1g_0\right]f_R
\nonumber \\
&-&6 \bar{c}\lambda_0a\dot{a}.          \label{1.9}
\enea
This equation may be written as
\bea \label{0.8'''}
f_{RR}\dot{R}&=&-\frac{\Theta}{6a^3\phi^2c_1 g_0}-2\frac{d}{dt}\ln(a\phi)f_R\\
\nonumber
&+&\frac{4B_0\dot{\phi}\omega(\phi)}{3\phi^2c_1g_0}
+3f_RH-\frac{\bar{c}\lambda_0H}{c_1g_0a\phi^2}.
\enea
The Friedmann equation is obtained by construction of the zero Hamiltonian constraint as
\bea \label{Ham}
H&=&\dot{a}\frac{\partial\mathcal{L}}{\partial\dot{a}}+\dot{R}\frac{\partial\mathcal{L}}{\partial\dot{R}}+
\dot{\phi}\frac{\partial\mathcal{L}}{\partial\dot{\phi}}-\mathcal{L}\\ \nonumber
&=&f+6f_{RR}\dot{R}H+6f_RH^2+12f_RH(\frac{\dot{\phi}}{\phi})\\ \nonumber
&-&4(\frac{\dot{\phi}}{\phi})^2\omega(\phi)-f_R(R-\frac{6k}{a^2})-\frac{V(\phi)}{{\phi}^2}=0,
\enea
or
\bea \label{Ham'}
H&=&f-\frac{\Theta H}{\phi^2 a^3 c_1 g_0}+12f_R H^2\\ \nonumber
&-&\frac{6\bar{c}\lambda_0 H^2}{c_1a\phi^2g_0}
-f_R(R-\frac{6k}{a^2})\\ \nonumber
&-&\frac{V(\phi)}{\phi^2}+\frac{\dot{\phi}}{\phi^2}\omega(\phi)\left(\frac{8H
B_0}{c_1g_0}-4{\dot{\phi}}\right)=0,
\enea
where we have used of (\ref{0.8'''}) and that
$$
\frac{\dot{\phi}}{\phi}=\frac{d}{dt}\ln(\phi)\,\,\,,\,\,\,\frac{d}{dt}\ln(\frac{a\phi}{\phi})=\frac{\dot{a}}{a}=H.
$$ 
\section{$f(R)$ Cosmological Models}
\label{5}

To find some $f(R)$ scalar-tensor cosmological models consistent with the Noether symmetry, we first rewrite the constraint equation (\ref{0.8'}) as follows
\bea\label{0.8''}
&&c_1 a^2(3a^4\phi^2 f-3a^4V(\phi)-12k a^2\phi^2 f_R)g_0\\ 
&+&Ba^6(2\phi f-\frac{dV}{d\phi})
=\bar{c}\lambda_0a^3(Ra^2-6k).\nonumber
\enea
Then, we study the different cases according to some specific choices for $
\omega(\phi), V(\phi), B_0, g_0, c_1$ and $\Theta$. In all cases, except one, we will assume $\bar{c}=0$.

\subsection{$B_0=V(\phi)=0$}

In this case, Eq.(\ref{0.8''}) is reduced to 
\bea\label{f_R0}
a^2\phi^2 f-a^2V(\phi)-4k\phi^2 f_R=0,
\enea
or
\bea\label{f_R}
f_R=\frac{a^2}{4k}f.
\enea
Using (\ref{f_R}) in the Friedmann equation (\ref{Ham'}), we obtain
\bea \label{Ham1}
f\left(\frac{5}{2}+\frac{3a^2H^2}{k}-\frac{Ra^2}{4k}\right)=
\frac{\Theta H}{c_1 g_0\phi^2 a^3}+4\omega(\phi)(\frac{\dot{\phi}}{\phi^2})^2_.
\enea

I) Imposing $\Theta=\omega(\phi)=0$ leads to ($f\neq0$)
\bea\label{I}
R=12H^2+\frac{10k}{a^2}.
\enea
Considering the general expression for the Ricci scalar 
\bea\label{Ricci}
R=12H^2+6a H H'+\frac{6k}{a^2},
\enea
one may find the following differential equation by equating the RHS of (\ref{I}) and (\ref{Ricci}) 
\bea\label{H(a)}
(H^2)'-\frac{4k}{3a^3}=0,
\enea
where $'$ denotes derivative with respect to $a$.
The differential equation (\ref{H(a)}) is simply solved as
\bea\label{H^2(a)}
H^2=-\frac{2k}{3a^2}+d_1,
\enea
where $d_1$ is the integration constant. This equation determines the cosmological
dynamics. Now, we obtain $f(R)$. To this end, we insert (\ref{H^2(a)}) into
(\ref{I}) to obtain $a(R)$. Then, we calculate $f_R$ by using $a(R)$ and
$f_R=\frac{df}{da}\frac{da}{dR}$ as follows
\bea\label{f(a)?}
f_R=-\frac{df}{da}\frac{a^3}{4k}.
\enea
Finally, using (\ref{f(a)?}) in (\ref{f_R}) results in
\bea\label{f(a)}
f(a)=\frac{d_2}{a},
\enea
where $d_2$ is another integration constant. Putting $H^2$ from (\ref{H^2(a)})
in (\ref{I}), we obtain
\bea\label{R(a)}
R=\frac{2k}{a^2}+12d_1. 
\enea
Using this equation, we may transform $f(a)$ into
$f(R)$ as 
\bea\label{f(R)}
f(R)=d_2\left(\frac{R-12 d_1}{2k}\right)^{1/2}. 
\enea
This is viable for closed and open universes, $k=\pm1$. For the flat universe, we may take $d_2=\sqrt{2k}$ so that
\bea\label{k=0}
f(R)=({R-12 d_1})^{1/2}. 
\enea

II) Imposing $\Theta\neq0, \omega(\phi)=0$ in (\ref{Ham1}), results in
\bea \label{Ham2}
f=\frac{2k\Theta H}{c_1g_0\phi^2a^3(2k-3a^3HH')},
\enea
where use has been made of (\ref{Ricci}). Note that $f_R=\frac{\partial
f}{\partial a}\frac{da}{d R}$, so we may calculate separately $\frac{\partial f}{\partial a}$ and
$\frac{da}{d R}$ using (\ref{Ham2}) and (\ref{Ricci}), respectively. Then,
(\ref{f_R}) casts into a differential equation for $H^2$ as
\bea \label{Diff}
(H^2)''=(H^2)'\left(\frac{2k}{3H^2a^3}-\frac{5}{a}\right)-\frac{16k^2}{9a^6}\frac{1}{(H^2)'}+\frac{8k}{3a^4}.
\enea
By solving this differential equation we may find $H(a)$ which determines the cosmological dynamics. Moreover, we may insert $H(a)$ into (\ref{Ricci}) to find $a(R)$. Then, we can insert both $H(a)$ and $a(R)$ into (\ref{Ham2})
to obtain $f(R)$.

\subsection{$B_0=0, V(\phi)=\frac{1}{2}m^2 \phi^2$}

In this case, Eq.(\ref{f_R0}) leads to
\bea\label{f_R1}
f_R=\frac{a^2}{4k}f-\frac{m^2 a^2}{8k}.
\enea
By inserting $f_R$ from (\ref{f_R1}) into the Friedmann equation (\ref{Ham}), we obtain
\bea \label{Ham3}
&&({10k+12 a^2 H^2-a^2R})f=
\frac{4k\Theta H}{c_1 g_0\phi^2 a^3}\\ \nonumber
&+&{5m^2k+2m^2a^2(3H^2-R/4)}
+16k\omega(\phi)(\frac{\dot{\phi}}{\phi^2})^2.
\enea

I) Imposing $\Theta=\omega(\phi)=0$ in (\ref{Ham3}) leads to
\bea\label{Ii}
f=\frac{5k+6a^2H^2-a^2R/2}{10k+12a^2H^2-a^2R}m^2.
\enea
By inserting $R$ from (\ref{Ricci}) into (\ref{Ii}) we obtain
\bea\label{Ii1}
f=\frac{1}{2}m^2.
\enea
Obviously, this case is not physically viable because it leads to a constant
$f$ with no cosmological solutions.     

II) Imposing $\Theta\neq0, \omega(\phi)=0$ in (\ref{Ham3}), and inserting $R$ from (\ref{Ricci}) results in
\bea\label{Ii2}
f=\frac{1}{2}m^2+\frac{2k\Theta H}{c_1g_0\phi^2a^3(2k-3a^3HH')}.
\enea
Using $f_R=\frac{\partial f}{\partial a}\frac{da}{d R}$, (\ref{Ricci}) and (\ref{Ii2}) we
obtain the same differential equation for $H^2$ as (\ref{Diff}).
By solving this equation for $H(a)$, the cosmological dynamics is obtained. By inserting $H(a)$ into (\ref{Ricci}) we find $a(R)$, and by using $H(a)$ and $a(R)$ in (\ref{Ii2}) we obtain $f(R, \phi)$. Now, in order to remove $\phi$ in favor of $R$ within $f(R, \phi)$, we first rewrite Eq.(\ref{0.1}) as follows
\beq
3c_1 g_0 a\omega(\phi)+B_0a\frac{d\omega}{d\phi}=0,   \label{0.1'}
\eneq
where we have used of (\ref{1.66}). This equation allows us to obtain $\phi$
in terms of $a$ as a function $\phi(a)$. On the other hand, we have $a(R)$
from (\ref{Ricci}). Therefore, combining $\phi(a)$ and $a(R)$ we may obtain $\phi(R)$
by which we can replace $\phi$ in $f(R, \phi)$ in terms of $R$ and obtain
the desired $f(R)$.

\subsection{$B_0\neq0, V(\phi)=0$}

In this case, Eq.(\ref{0.8''}) is reduced to 
\bea\label{f_R2}
f_R=\frac{(3c_1g_0\phi+2B_0)a^2}{12kc_1g_0\phi}f.
\enea
Inserting $f_R$ from (\ref{f_R2}) into the Friedmann equation (\ref{Ham}),
and using the following definitions
\begin{equation}\label{def}
\left \{ \begin{array}{ll} \alpha=30kc_1g_0\phi+12k B_0,
\\
\beta=12kc_1g_0\phi,
\\
\xi=36c_1g_0\phi+24B_0=12\gamma,
\\
\gamma=3c_1g_0\phi+2B_0,
\\
\Delta=c_1g_0\phi^2,
\\
\zeta=\alpha-6k\gamma
\\
\mu=4(\frac{\dot{\phi}}{\phi})^2,
\\
\nu={8B_0\dot{\phi}}/{c_1g_0\phi^2},
\end{array}\right.
\end{equation}
together with (\ref{Ricci}), we obtain 
\bea\label{f}
f\left(\frac{\zeta-6\gamma a^3 H H'}{\beta}\right)=\frac{\Theta H}{\Delta a^3}+(\mu-\nu
H)\omega(\phi).
\enea

I) Imposing $\Theta=0, \omega(\phi)\neq0$ in (\ref{f}) leads to
\bea\label{f1}
f=\frac{D-EH}{G-Ma^3H H'},
\enea
where
\begin{equation}\label{def1}
\left \{ \begin{array}{ll} 
D=\beta\mu\omega(\phi),
\\
E=\beta\nu\omega(\phi),
\\
G=\zeta=\beta,
\\
M=6\gamma.
\end{array}\right.
\end{equation}
Using $f_R=\frac{\partial f}{\partial a}\frac{da}{d R}$, (\ref{Ricci}), (\ref{0.1'}) and (\ref{f1}), we obtain a differential equation for $H^2$.
By solving this equation for $H(a)$ we obtain the cosmological dynamics. Inserting $H(a)$ into (\ref{Ricci}) results in $a(R)$. Then, we may use $H(a)$ and $a(R)$ in (\ref{f}), use Eq.(\ref{0.1'}) and $a(R)$ to obtain $\phi(R)$, and finally obtain $f(R)$.

II) Imposing $\Theta\neq0, \omega(\phi)\neq0$ in (\ref{f}) leads to
\bea\label{f3}
f=\frac{a^3(\bar{D}-\bar{E} H)+\bar{F} H}{a^3(\bar{G}-a^3\bar{M} H H')},
\enea
where
\begin{equation}\label{def2}
\left \{ \begin{array}{ll} 
\bar{D}=\Delta\beta\mu\omega(\phi),
\\
\bar{E}=\Delta\beta\nu\omega(\phi),
\\
\bar{F}=\beta\Theta,
\\
\bar{G}=\beta\Delta,
\\
\bar{M}=6\gamma\Delta.
\end{array}\right.
\end{equation}
Using $f_R=\frac{\partial f}{\partial a}\frac{da}{d R}$, (\ref{Ricci}), (\ref{0.1'}) and (\ref{f3}), we obtain a differential equation for $H^2$.
By solving this equation we obtain the cosmological dynamics $H(a)$. 
Moreover, similar to the previous case we may obtain $f(R)$

\subsection{$\bar{c}\neq0, V(\phi)=g_0=0$}

In this case, Eq.(\ref{C}) reads as
\beq\label{C1}
C=-\frac{2B}{\phi}\frac{f_R}{f_{RR}}+\frac{\bar{c}\lambda_0}{a\phi^2 f_{RR}}.
\eneq
Inserting $C$ from (\ref{C1}) into (\ref{0.8}), results in
\bea\label{1.8}
(2B_0a^3\phi)f=a^3 B_0 \frac{dV}{d\phi}+\bar{c}\lambda_0(Ra^2-6k).
\enea
Unlike the previous procedure, in this case the constraint equation is not a differential equation containing $f_R$. To find an equation containing
$f_R$ we use (\ref{Ham'}). We evaluate $f_{RR}\dot{R}$ from (\ref{1.8})
and put it in (\ref{Ham}) to obtain
\bea \label{Ham4}
&f&+\frac{3\bar{c}\dot{\lambda}(\phi)}{a B_0}H-\frac{3\bar{c}{\lambda(\phi)}}{a
B_0}H^2+6f_RH^2\\ \nonumber
&+&12f_RH(\frac{\dot{\phi}}{\phi})
-4(\frac{\dot{\phi}}{\phi})^2\omega(\phi)-f_R(R-\frac{6k}{a^2})-\frac{V(\phi)}{{\phi}^2}=0.
\enea
Since $g_0=0$, we have $\omega(\phi)=1$. Now, we insert (\ref{1.8}) and (\ref{Ricci})
into (\ref{Ham4}) and obtain the required differential equation containing
$f_R$.
\bea \label{Ham5}
&f_R&[-6H^2+12H(\frac{\dot{\phi}}{\phi})-6a H H']\\ \nonumber
&=&-\frac{3\bar{c}{\lambda(\phi)}}{a
B_0}[H^2+a H H'+H\frac{\dot{\lambda}}{\lambda}]\\ \nonumber
&+&4(\frac{\dot{\phi}}{\phi})^2-\frac{1}{2\phi}\frac{dV}{d\phi}
+\frac{V(\phi)}{{\phi}^2}.
\enea
For simplicity we consider $V(\phi)=0$, and obtain
\bea \label{Ham6}
f_R=\frac{\frac{3\bar{c}{\lambda(\phi)}}{a
B_0}[H^2+a H H'+H\frac{\dot{\lambda}}{\lambda}]-4(\frac{\dot{\phi}}{\phi})^2}{6H^2-12H(\frac{\dot{\phi}}{\phi})+6a H H'}.
\enea
This equation may be solved for $f(R)$ by using $H(a)$, $a(R)$, and $\phi(R)$. Using $f_R=\frac{\partial f}{\partial a}\frac{da}{d R}$, (\ref{Ricci}) and (\ref{1.8}),  a differential equation for $H^2$ is obtained as
\bea \label{Ham7}
&&(H^2)''=\frac{1}{Ha^3(2\alpha\frac{\dot{\phi}}{\phi}+\alpha \mu)-a^4\beta}\times
\\ \nonumber
&&[a^2(H^2)'[3\alpha(H^2)+\frac{\alpha
a}{2}(H^2)'+(8\alpha\frac{\dot{\phi}}{\phi}+5\alpha\mu)H-5\beta a]+\\ \nonumber
&&4\alpha a(H^2)[(H^2)-2H\frac{\dot{\phi}}{\phi}]+4\alpha k[\frac{(H^2)'}{2}-\frac{H^2}{a}-\mu\frac{H}{a}]+4\beta
k],
\enea
where
\begin{equation}\nonumber
\left \{ \begin{array}{ll} 
\alpha=\frac{3\bar{c}\lambda(\phi)}{B_0},
\\
\beta=4(\frac{\dot{\phi}}{\phi})^2,
\\
\mu=\frac{\dot{\lambda}}{\lambda}.
\end{array}\right.
\end{equation}
This equation may be solved for $H^2(a)$ by using $\phi(a)$.

\vspace{2mm}

\section{Conclusions}
\label{7}
In this paper, we have investigated the conditions for the existence of Noether symmetry in a $f(R)$ scalar-tensor theory of gravity in which the
Ricci function $f(R)$, the scalar field potential $V(\phi)$ and the coupling function $\omega(\phi)$ are generally unknown.
We have shown that the Noether symmetry may exist and further obtained a constraint between $f(R)$, $V(\phi)$ and $\omega(\phi)$. For specific choices of the functions $\omega(\phi), V(\phi)$, the parameters $B_0, g_0, c_1,$ and the constant of motion $\Theta$, we have obtained explicitly the functions $f(R)$ and $H(a)$. For other cases, we have found the corresponding differential equations which can only be solved numerically.


\section*{Acknowledgment}

This work has been supported financially by Research Institute
for Astronomy and Astrophysics of Maragha (RIAAM) under research project
NO.1/2077.

\end{document}